# Battery-free, stretchable, and autonomous smart packaging

Ali Douaki [a e f*], Mukhtar Ahmed [a], Edoardo Longo [b], Giulia Windisch [b], Raheel Riaz [a], Sarwar Inam [a], Thi Nga Tran [c], Evie L. Papadopoulou [d], Athanassia Athanassiou [c], Emanuele Boselli [b], Luisa Petti [a *], Paolo Lugli [a *]

[a] *Sensing Technologies Laboratory (STL), Faculty of Engineering, Free University of Bozen-Bolzano, Piazza Università 5, 39100, Bozen, Italy.*

[b] *Faculty of Agricultural, Environmental and Food Sciences, Free University of Bozen-Bolzano, Piazza Università 5, 39100, Bozen, Italy.*

[c] *Smart Materials Group, Istituto italiano di Tecnologia, via Morego 30, 16163 Genova, Italy*

[d] BeDimensional S.p.A. *Lungotorrente Secca 30 R, 16163 Genova, Italy*

[e] *Optoelectronics Research Line, Istituto italiano di Tecnologia, via Morego 30, 16163 Genova, Italy*

[f] *Dip. di Scienze e Metodi dell'Ingegneria, Università di Modena e Reggio Emilia, via Amendola 2, 42122 Reggio Emilia, Italy.*

**Abstract**

In the food industry, innovative packaging solutions are increasingly important for reducing food waste and for contributing to global sustainability efforts. However, current food packaging is generally passive and unable to adapt to changes in the food environment in real-time. To address this, we have developed a battery-less and autonomous smart packaging system that wirelessly powers closed-loop sensing and release of active compounds. This system integrates a gas sensor for real-time food monitoring, a Near-Field Communication (NFC) antenna, and a controlled release of active compounds to prevent quality deterioration in the complex food environment. We have demonstrated the ability of the developed smart packaging system, to continuously monitor the freshness of fish products and to trigger the release of active compounds when the food starts to spoil. The system was able to extend the shelf-life of the food product up to 14 days, due to the controlled release of antioxidant and antibacterial compounds. Our system could pave the way towards an Internet of Things solution that addresses protection, active prevention of food spoilage and sustainability, facing all the current challenges of the food packaging industry.

# Introduction

The need for sustainable and safe packaging solutions continues to grow, driven by increasing consumer awareness and environmental concerns [1]. This has become even more crucial given the persistent global issues of food waste and spoilage. According to estimates, roughly one-third of all food produced is wasted or lost annually, leading to significant economic, social, and environmental impacts [2]. In addition to the financial cost of wasted food, which can amount to billions of dollars annually, the production and transportation of food that is ultimately discarded also have environmental consequences, including greenhouse gas emissions and resource depletion [3]. The main cause of food spoilage is oxidation due to contact with air over a prolonged period and, secondly, the uncontrolled development of micro-organisms [4,5]. To prolong the shelf life of food, considerable efforts are being undertaken to delay the occurrence and extent of these processes using advanced food packaging. It is therefore crucial to develop sustainable and innovative packaging solutions that can preserve the quality and extend the shelf life of products, ultimately reducing waste and benefiting both consumers and the environment.

Advancements in active and intelligent packaging technologies have been undertaken to reduce food waste, primarily through shelf-life extension or by monitoring food product quality. Active packaging incorporates components like oxygen scavengers [6,7], moisture absorbers [6,8], and antimicrobial agents [9–13], to enhance product quality and prolong shelf life [14,15]. On the other hand, intelligent packaging employs sensors, indicators, and RFID tags for real-time monitoring of product conditions [16–21]. However, both technologies have limitations that render them ineffective, especially when applied independently.

The existing working mechanism of innovative packaging for extending food shelf life relies on oxygen absorption, bio-active compound release, and volatile organic compound (VOC) concentration monitoring, which are only moderately effective [22]. This is because these technologies generally exhibit a passive nature, lacking the ability to actively adapt to fluctuations in the food package environment [22]. For instance, active packaging, while autonomously functional, is incapable of responding to changing environmental variables, and the continuous release of bio-active compounds into the packaging space, governed by Fick's law of diffusion, can compromise the food's freshness [23]. This may even lead to microbial resistance in food products, posing potential public health risks and economic losses [24,25]. Efforts to synchronize the release rate of these compounds with the spoilage rate of food products have proven challenging and largely unsuccessful [26]. Intelligent packaging, on the other hand, depends on external sensors and indicators to provide real-time data on product conditions. However, this technology falls short in applications requiring a longer shelf life, as it does not actively interact with the product environment, hence, a quest to find new smart packaging solutions has emerged. Moreover, so far, multiple separate devices have been reported in the literature such as gas sensors [27], NFC antennas, and controlled release mechanisms [9]. A smart packaging solution that combines the strengths of active and intelligent packaging, while addressing their limitations, has the potential to significantly improve food safety, extend shelf life, and reduce waste. This would benefit consumers, manufacturers, and the environment.

To achieve an improved food packaging system, it should meet the following requirements: (i) controlled release of antioxidants with zero leakage in the off state, (ii) integration of sensing and autonomous, closed-loop, food freshness management, (iii) wirelessly powered to avoid having batteries in direct contact with food, (iv) stability under harsh environment. Such system should also fulfill the 4S (scalable, stretchable, self-powered, and sensing).

This study aims to advance the existing literature by developing a comprehensive smart packaging system through the integration of individual devices into a complete battery-free, stretchable, and autonomous smart packaging, designed to extend the shelf life of food products. The system combines the benefits of both active and intelligent packaging technologies, with the intelligent packaging component monitoring the freshness of the food product and activating the release of the antioxidants from the active packaging component only when the food starts to deteriorate. To enable autonomous operation, the two components are linked via an NFC antenna that wirelessly powers the system and triggers the release of the antioxidants. Each component of the smart packaging system was characterized separately and then as a complete system. To validate the efficacy of the smart packaging system, we employed gas chromatography/ mass spectrometry (GC/MS) to measure the release of cinnamaldehyde (CA) and eugenol (EG) inside the packaging space. Our results show that food waste can be reduced by using an innovative smart packaging system. Furthermore, the system can also be integrated with IoT to improve supply chain management.

## Results and discussion

### Wireless and battery-free smart packaging system

The developed smart packaging utilizes a gas sensor, an NFC antenna, and smart materials to control the release of active compounds. Figure 1a shows a schematic illustration of the smart packaging, highlighting the three main components, i.e. the single-walled carbon nanotube (SWCNT) gas sensor, the wireless platform, and the controlled release system. The wireless platform consists of (i) a power harvesting coil that operates by magnetic inductive coupling at a resonance frequency of 13.56 MHz, to power the system, (ii) a near-field communication (NFC) system on chip (M24LR16E) that supports wireless communication, (iii) a SWCNT gas sensor that serves as a food spoilage detector and the switch to trigger the release of the antimicrobial agent, and (iv) electrospun mat containing active compound that releases under a thermal stimulation provided by the heater at the bottom of the mat "Poly(3,4-ethylenedioxythiophene)-poly(styrenesulfonate) (PEDOT:PSS)" (Figure 1b).

Figure 1c elucidates the working principle of smart packaging, which draws inspiration from the functional mechanism of a neuron. Like a neuron's response to a chemical stimulus, the smart packaging transduces chemical signals into electrical impulses. In this case, food spoilage coincides with the formation of ammonia ($NH_3$) used here as an indicator — being released into the headspace of the packaging. Ammonia interacts with the SWCNT gas sensor, causing an increase in its resistance. This resistance change affects the connected antenna's resonance frequency, enhancing the NFC antenna's gain and, consequently, the voltage harvested by the NFC antenna. This harvested voltage is then supplied to the PEDOT:PSS layer to elevate the mat's temperature above the lower critical solution temperature (LCST) of 32°C for Poly(N-isopropylacrylamide) (PNIPAM). The latter is used to control active compound release. At room temperature, PNIPAM remains in a swelled state, inhibiting the release of the active compounds (Note 1. in SI). However, surpassing the LCST causes PNIPAM to collapse, triggering the discharge of the active compound from the fibers [9]. The developed packaging works as a closed-loop system that continuously monitors and adapts the conditions of the food packaging headspace, and thus preserves in a dynamic way the food freshness (Figure 1d).

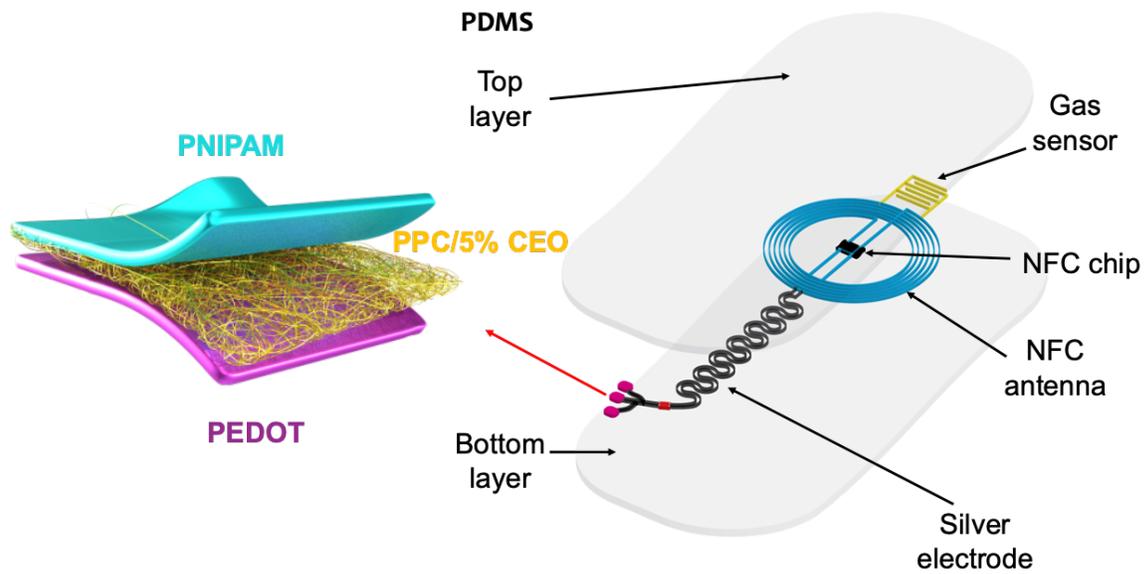

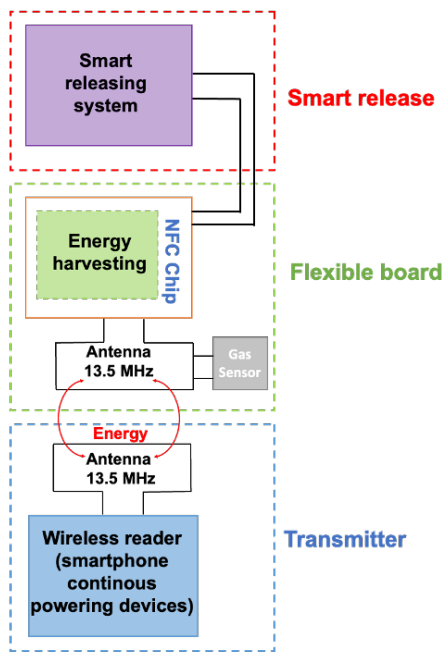

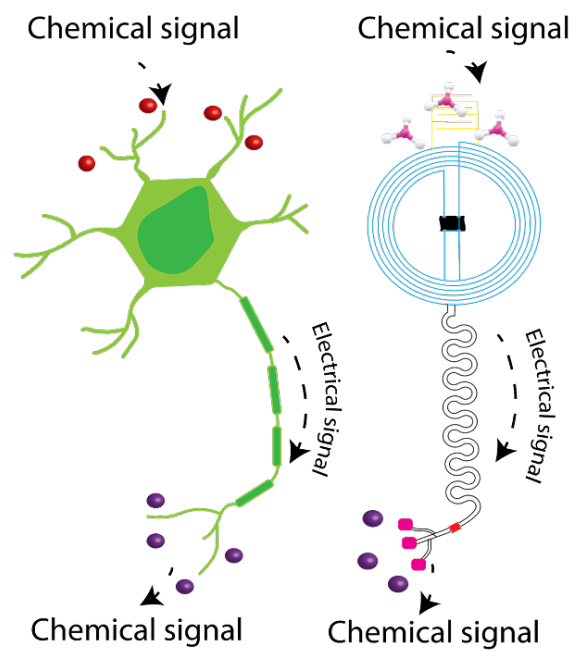

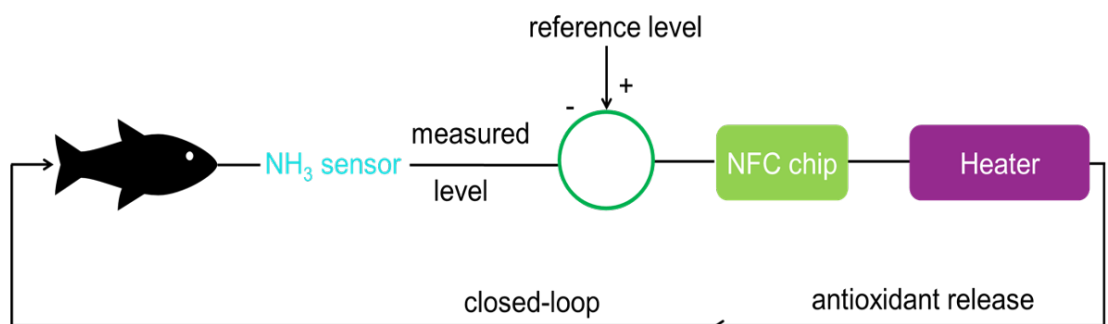

*Figure* **1.** ***Device concept.*** *(**a**) Schematic illustrating the exploded view of the complete hybrid, battery-free system. (**b**) different parts of the smart packaging. (**c**) Working mechanism of the smart packaging. (**d**) Image illustrating the closed-loop system.*

**Characterization of the gas sensor**

Previous studies have established ammonia ($NH_3$) as a robust biomarker for detecting fish spoilage [28]. This is because $NH_3$ is produced by bacteria during protein spoilage and accumulates in the headspace of spoiled fish packaging. Consequently, $NH_3$ detection offers a non-invasive alternative to conventional methods and the capability for a rapid and real-time monitoring [29–32]. In this work, we have developed the SWCNT gas sensor to assess the freshness of fish products by quantifying $NH_3$ levels in the headspace of the food packaging. This section focuses on the fabrication and characterization of this gas sensor. Figure 2a displays a scanning electron microscope (SEM) image of a 25-layer spray-deposited SWCNT film. This film was optimized in terms of number of sprayed layered to attain an initial resistance of 1 kOhm (explained in section 5). The image reveals a uniform, homogeneous, and sparsely distributed network of SWCNTs. Figures 2b, 2c, and 2d illustrate the SWCNT gas sensor's resistance variation over time following exposure to different $NH_3$, $CH_4$, and $CO_2$ gas concentrations. Additionally, Figure 2e exhibits the gas sensor's linear response to $NH_3$ increasing concentrations ranging from 15 ppm to 90 ppm, followed by sensor saturation. Notably, the SWCNT gas sensor demonstrated heightened sensitivity to $NH_3$ compared to $CH_4$ and $CO_2$. The resistance increase is attributed to the SWCNT gas sensor's mechanism, extensively documented in literature [33–36]. As the smart packaging developed in this research utilizes the resistance increase in the gas sensor to initiate release mechanisms. Therefore, gas selectivity is a crucial aspect for ensuring reliable and precise detection, as lower specificity for gases could lead to false system activation. Consequently, the selectivity of the gas sensors against $NH_3$, $CH_4$, and $CO_2$ was thoroughly examined. The results demonstrated a significant 13% response increase in the gas sensor when exposed to 90 ppm of $NH_3$. In contrast, exposure to 500 ppm of $CH_4$ and 7500 ppm of $CO_2$ resulted in only a 5% and 3.5% response increase, respectively, as depicted in Figure 2f. Despite higher concentrations of $CH_4$ and $CO_2$, the sensor's response to $NH_3$ was more pronounced, indicating the sensor's favorable selectivity towards $NH_3$. Afterwards, the change of SWCNT gas sensor in presence of $NH_3$ without recovering the gas sensor was investigated and reported in Figure 2g, as we can observe the increase in $NH_3$ concertation led to an increase in the resistance of the gas sensor from 900 Ohm to 1800 Ohm in a concentration of $NH_3$ ranging from 15 ppm to 90 ppm. This increase of resistance will later be employed to trigger the release of the active compounds.

Moreover, the humidity notably influences the resistance of SWCNTs as has been previously reported[37]. Given that humidity levels inside food packaging often hover around 90% (Figure S1), this becomes a significant factor in gas sensor design for such applications. To address humidity interference, a thin polydimethylsiloxane (PDMS) layer was applied to passivate the SWCNT gas sensor. PDMS, being porous yet hydrophobic, permits gas molecules to permeate while barring water molecules [38]. Therefore, the influence of PDMS passivation on the sensor's sensitivity was assessed, as shown in Figure 2h. The figure displays the SWCNT gas sensor's response to 90 ppm of $NH_3$. The passivated sensor showed a mere 5% decrease in response, likely due to the $NH_3$ molecules' ability to penetrate the PDMS layer easily. Mechanical stability of the gas sensor is another important parameter, hence, the effect of mechanical stress in terms of bending was investigated. As shown in Figure 2i, the gas sensor showed a good stability even after 5000 bending cycles where the gas sensor lost around 5% of the original sensitivity.

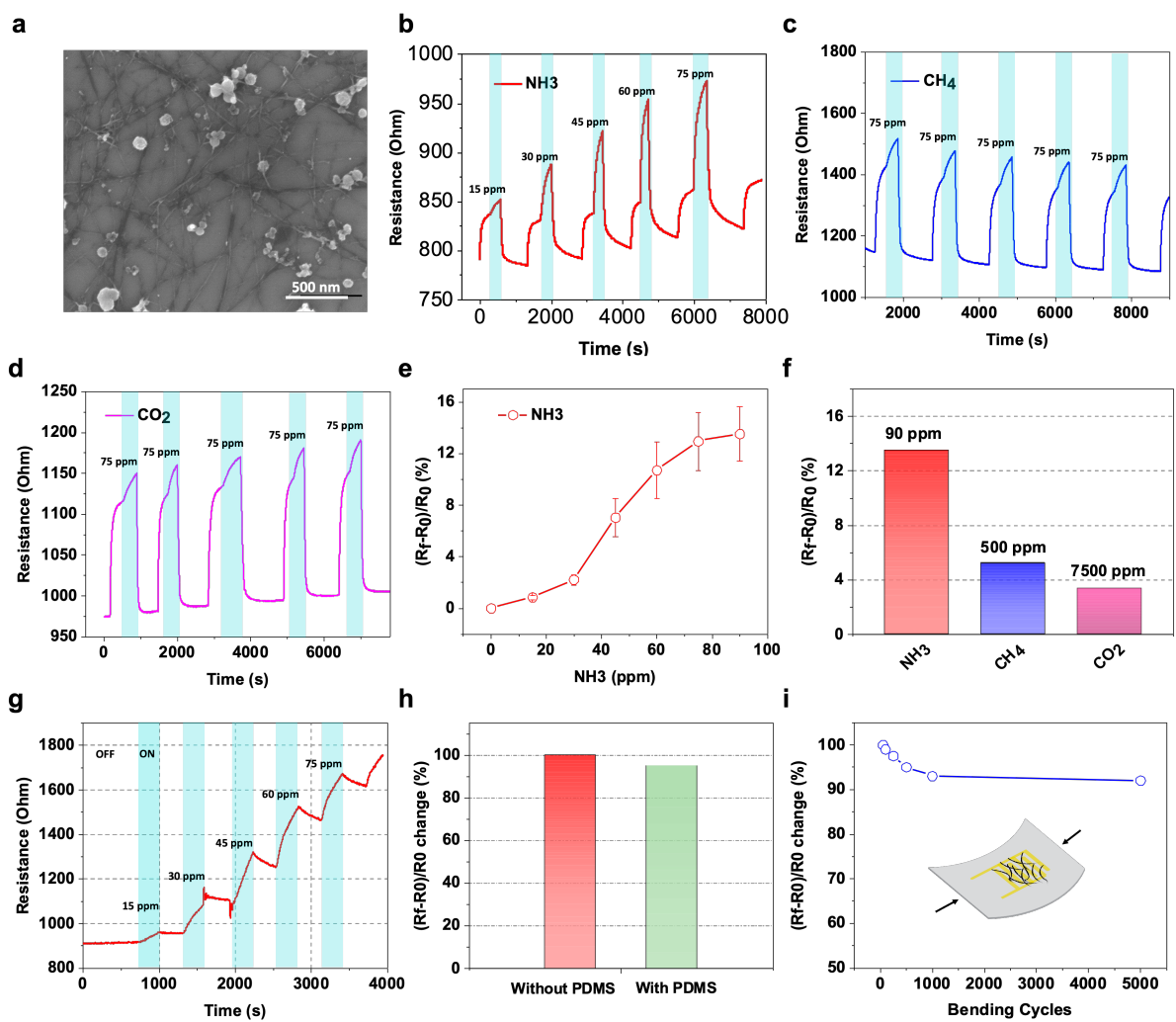

*Figure 2. (a) Scanning electron microscopy (SEM) image of the spray-coated SWCNTs on a silicon substrate. (b) Response of CNTs gas sensor towards ammonia for concentrations ranging from 15 to 75 ppm with a step of 15 ppm. (c) Response of CNTs gas sensor towards methane. (d) Response of CNTs gas sensor towards carbon dioxide, (e) Calibration curve of CNTs gas sensor towards ammonia. (f) Selectivity test of CNTs gas sensor. (g) Resistance change of the gas sensor in presence of different ammonia concentrations ranging from 15 to 90 ppm with a step of 15 ppm. (i) The response of the CNTs gas sensor towards ammonia without and with PDMS for 90 ppm of $NH_3$. (j) The effect of the mechanical stress "bending" on the performance of the CNTs gas sensor om PDMS substrate "in presence of 90 ppm $NH_3$".*

## NFC antenna in smart food packaging

The third part of the smart packaging is the NFC antenna which is the skeleton of smart packaging. The NFC antenna was employed mainly for two reasons, first to connect both the gas sensor (intelligent packaging) and the releasing mat (active packaging) and hence, to trigger the release of CEO once the food starts to get spoiled. Secondly, to wirelessly harvest energy and power up the heater integrated inside the mat. Figure 3a illustrates the schematic layout of the NFC antenna along with the chip. Following the fabrication process, the NFC was characterized, as depicted in Figure 3b. The antenna exhibited its peak resonance at approximately 14 MHz, while the bandwidth (3 MHz) remained within the standard NFC 13.56 MHz frequency range.

Given that food products are subjected to significant mechanical stress during transportation and shelf storage, it is imperative to investigate the mechanical stability and robustness of the antenna, especially considering its susceptibility to such stresses. Subsequent characterization was conducted after subjecting the antenna and the electrodes to various conditions, including different strain levels (ranging from 0% to 40%), various bending cycles (ranging from 0 to 5000), fluctuations in temperature (ranging from 5°C to 25°C), and changes in relative humidity (ranging from 20% RH to 80% RH), as illustrated in Figure 3 c-i. The overall resistance of the antenna conductive traces changed from 0.3 to 2.2 ohms, and the peak resonance frequency altered from 14 to 12 MHz, due to strain variance as depicted in Figure 3 c, d, and s3. Likewise, under the stress of 5000 bending cycles, the resonance frequency transitioned from 14 to 15.5 MHz, as depicted in Figure 3e. Figure 3f and 3.g illustrate the effects of temperature (changing from 14 MHz to 14.9 MHz) and humidity (shifting from 14 MHz to 15.5 MHz) on the antenna's resonance frequency. It is important to emphasize that although variations in the antenna resonance frequency were observed under different conditions, the antenna's bandwidth consistently remained within the standard NFC 13.56 MHz range (due to 3 MHz). on the other hand, the electrodes used to transport the harvested voltage from the NFC antenna showed a. slight increase on the resistance from 3 Ohm to 8 Ohm as shown in Figure 3h and 3i.

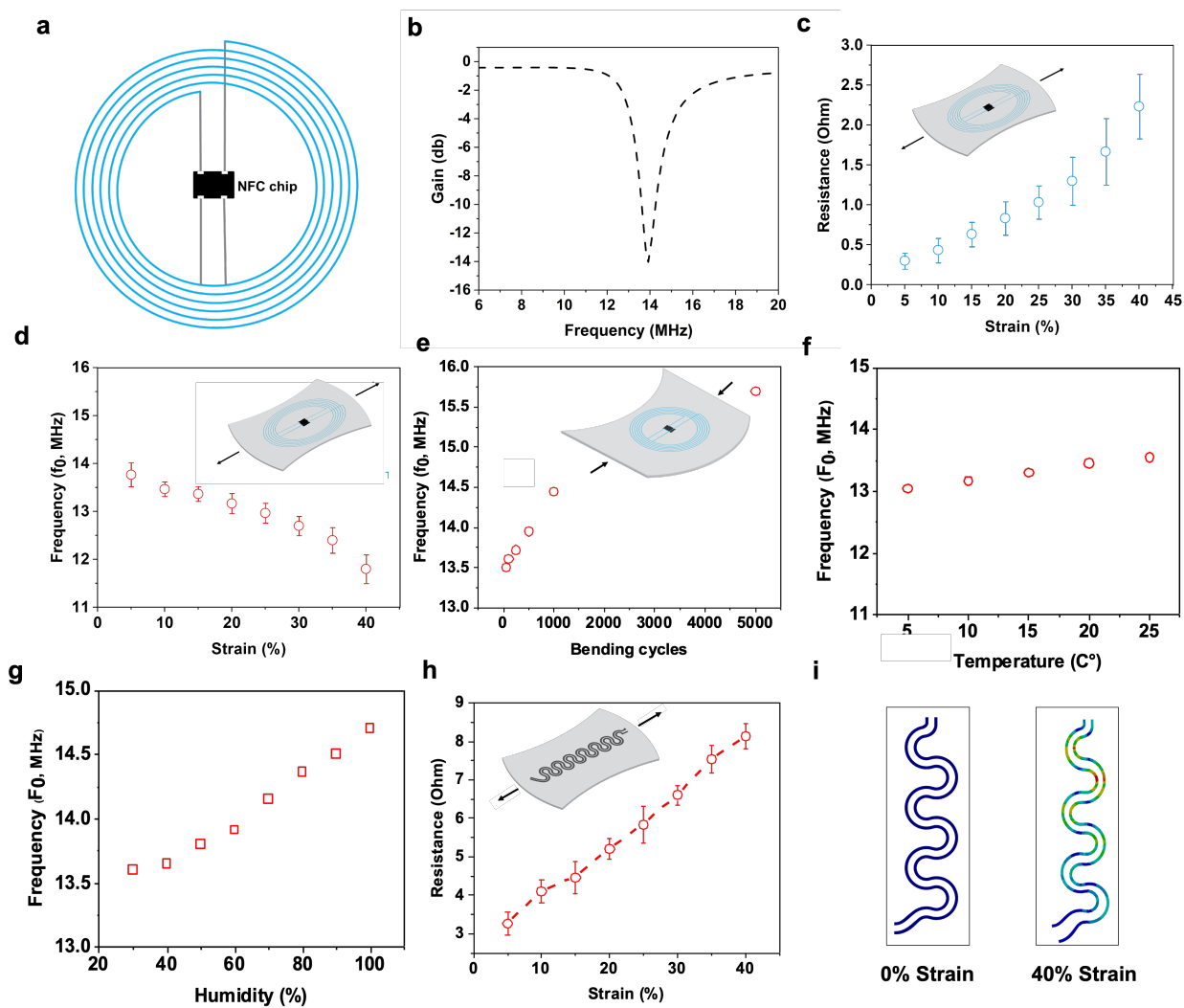

*Figure 3. (a)* an illustration of the NFC antenna with a chip. *(b)* The resonance frequency and the gain of the fabricated antenna. *(c)* The change in the resistance of the antenna coil under a mechanical strain. *(d)* The effect of mechanical stress on resonance frequency of the antenna. *(e)* Bending effect on the resonance frequency of the antenna. *(f)* The effect of temperature change on the resonance frequency of the antenna. *(g)* The effect of humidity change on the resonance frequency of the antenna. *(h)* Strains effect on the resistance of the electrodes, *(i)* ANSYS mechanical simulation of the antenna under different mechanical stress.

## Characterization of the smart packaging

Figure 4a shows an illustration of the final device "smart packaging" after the integration of intelligent and active packaging via the NFC. Figure 4b and Figure 4s show the setup used to determine the effects of ammonia on the gain of the NFC antenna. As the smart packaging will be implemented inside food packaging where the temperature and the humidity vary, it was important to investigate the effect of both variables on the gain of the NFC antenna.

Figure 4c and 5d illustrate the impact of varying temperature and humidity conditions on the antenna gain in both OFF and ON states of the smart packaging. Remarkably, the device demonstrates robust stability under humid conditions, attributed primarily to the protective layer of

PDMS. This layer acts as an effective moisture barrier due to its inherent hydrophobic properties. Furthermore, the device exhibited excellent stability across a spectrum of temperatures, ranging from 4°C, representative of refrigeration conditions, to 25°C, indicative of room temperature environments.

To simulate the spoilage of fish products and to study the effect of the latter on power harvesting of the NFC antenna in a controlled environment, a gas chamber was used to inject different concentrations of ammonia ranging from 5 ppm to 90 ppm. Figure 4e shows how different gases such as $NH_3$, $CH_4$, and $CO_2$ affected the gain of the NFC antenna. According to the results observed, ammonia had a larger impact on gain, and when concentrations of $NH_3$ were increased, gain also increased from 0.4 db to -5.6 db under 90 ppm of $NH_3$. Furthermore, the effect of different gases on the gain of the NFC antenna was also examined. As can be seen from the results, an increase in the concentration of $CH_4$ and $CO_2$ also increased the gain, however, this increase was negligible compared to the one caused by $NH_3$, thus illustrating the selectivity of the gas sensor. Figure 4f depicts the variation in wirelessly harvested voltage in response to $NH_3$. As the latter concentration augmented from 5 ppm to 40 ppm, the interaction between $NH_3$ molecules and the SCNT gas sensor led to an increase of the sensor's resistance. This increased resistance, when incorporated in parallel to the NFC coil, altered the antenna's resonance frequency, bringing it closer to 13.6 MHz — the frequency at which the harvested voltage peaks — Consequently, an increase in wirelessly harvested voltage was observed, reaching 5.8V. This voltage level was sufficient to power up the integrated heater in the active packaging, thereby initiating the release mechanism. As a point of clarification, the concentration of $NH_3$ in this experiment was chosen to simulate the spoilage of a food product, in other words, an increase in the concentration of $NH_3$, which represents the spoilage of the fish, lead to an increase in the voltage harvested, which is then used as a switch to trigger the release of the antioxidant.

Another factor affecting the harvested voltage was the position of the smart packaging inside the Helmholtz Coil Set-Up on the three axes, so it was important to optimize the position of the device inside the set-up as shown in Figure 4b. The results showed that at the inner side of the edges of the box and a distance of 5 cm of the Helmholtz Coil gave the highest output (Figure 4g).

Figure 4h illustrates the relationship between gain and temperature increase in response to varying $NH_3$ concentrations. In the absence of $NH_3$, both the gain and the temperature of the active packaging remained at 0.2 dB and 20 °C, respectively. However, as the $NH_3$ concentration gradually increased, the gain of the NFC antenna increased, while the temperature remained unchanged. Furthermore, once the NH₃ concentration reached 40 ppm the gain dropped to approximately -5 dB, corresponding to a harvested voltage of 3 V, the temperature of the active packaging film was increased to 27 °C by harnessing the ohmic heating. Subsequently, when the NH₃ concentration reached 40 ppm, the harvested voltage increased to 5.8 V, and the temperature of the active packaging film rose to 37 °C, a critical threshold for triggering the release mechanism.

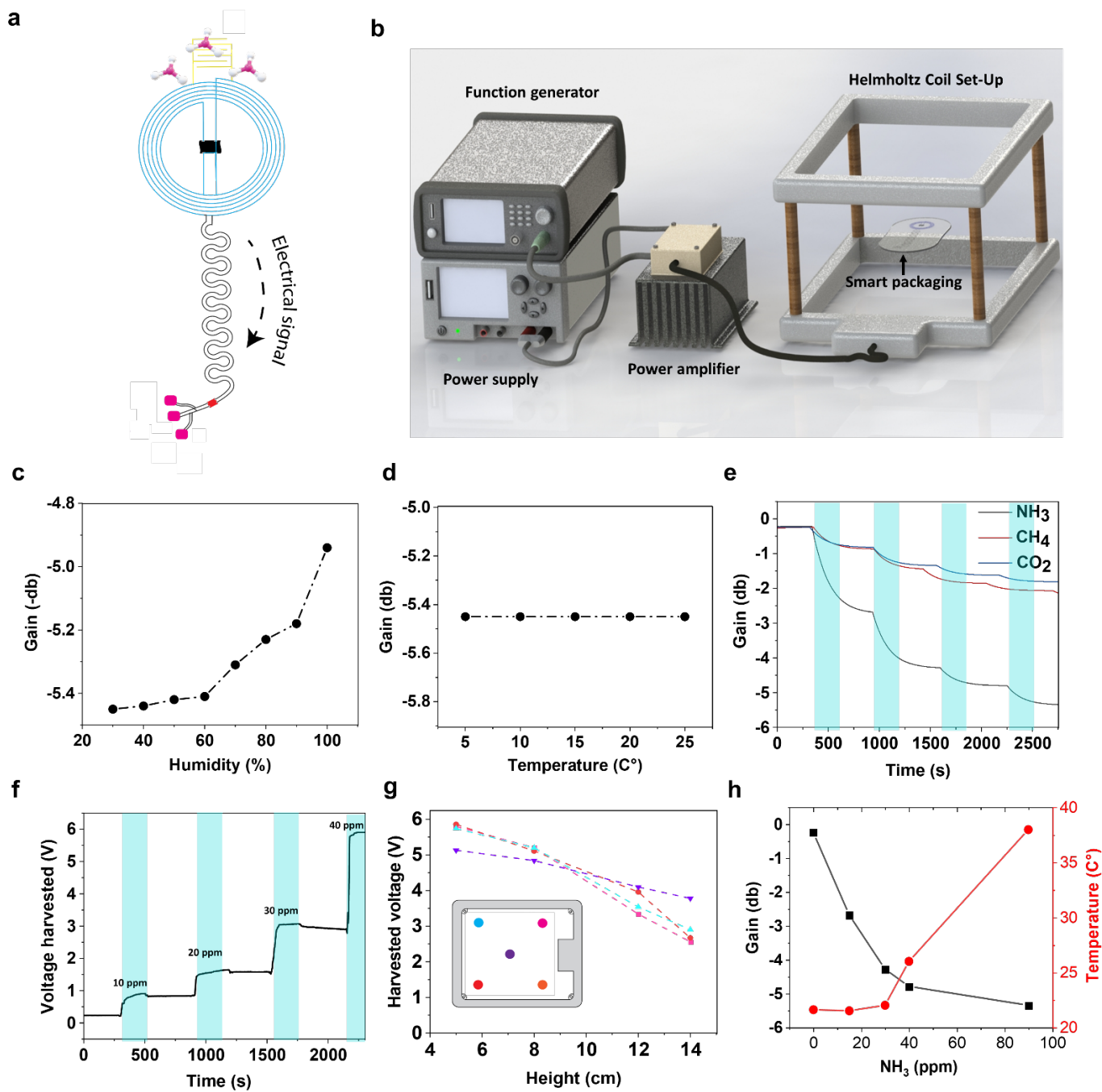

*Figure 4. (a)* An illustration of the final smart packaging composed of gas sensor, NFC antenna, and active packaging. *(b)* The experimental setup used to investigate the effect of different gases on the gain and harvested voltage from the NFC antenna. *(c)* The change in the gain of the antenna vs humidity. *(d)* The change in the gain of the antenna vs temperature. *(e)* The gain change of the NFC antenna in presence of NH₃, CH₄, and CO₂. *(f)* The change in the harvested voltage in presence of different NH₃ concentrations ranging from 5 ppm to 90 ppm. *(g)* The change in harvested voltage vs the position inside the Helmholtz Coil. *(h)* The effect of the

*active packaging size on the power needed to reach 35 °C. **(h)** The change in Gain and the temperature (active packaging film) vs different concentrations of NH₃.*

**Validation of the smart packaging**

The practical application of smart packaging was tested with fish in real conditions. Figure 5a illustrates the experimental setup, salmon and smart packaging were placed at the bottom and top of a box, respectively. Then, a headspace sample was extracted using an SPME fiber and analyzed by GC/MS. Total Volatile Basic Nitrogen (TVB-N) plays an important role in assessing the quality of meat products and is considered as a gold standard. When TVB-N reaches 25 mg per 100 g of salmon, deterioration is identified [39]. Hence, to correlate at which $NH^3$ concentration the salmon begins to deteriorate, TVB-N values were calculated while measuring NH3 concentration at room temperature using a commercial metal oxide gas sensor (Figure s7). At RT, the initial concentration of TVB-N was 1.3 ± 0.4 mg, which increased after nine hours of storage to 20.3 ± 0.01 mg. After 16 h of storage, the amount of TVB-N increased above the permitted limit across all samples with a corresponding NH3 concentration of 60 ppm. In light of this result, an NH3 threshold of 40 ppm was selected to trigger the release of the CEO.

The smart packaging system relies on CA and EG release to extend salmon shelf life. Monitoring the concentrations of these antioxidants and spoilage markers like 2-butanone and 3-methyl butanol was essential (Note 2. GC/MS In SI). Figure 5c and 6d show changes in spoilage markers and antioxidants in the headspace. Initially, at t=0, all markers were absent, indicating the fish's freshness. However, after 4 hours, concentrations of 2-butanone, cinnamaldehyde, and eugenol increased to 660 ppm, 270 ppm, and 70 ppm, respectively. The rise in 2-butanone signaled the beginning of spoilage, which in turn triggered the release of antioxidants, as evidenced by the presence of cinnamaldehyde and eugenol. After 24 hours, the concentrations of the markers decreased to 0 ppm, 45 ppm, 160 ppm, and 32 ppm for 2-butanone, 3-methyl butanol, cinnamaldehyde, and eugenol, respectively. These results demonstrate that fish spoilage was effectively prevented, as evidenced by the reduction in spoilage marker concentrations. However, the persistently high levels of antioxidants in the headspace suggest that not all cinnamaldehyde and eugenol were depleted, indicating the potential for longer fish preservation. Furthermore, Figure 5e reveals the cumulative release after 24h of CA and EG, with 67% and 100%, respectively, likely due to its lower concentration compared to cinnamaldehyde.

The smart packaging's performance was subsequently studied at 4 °C. The assessment at the lower temperature focused on measuring TVB-N levels to monitor fish freshness, instead of relying on VOCs. After four days of storage, the control sample (without smart packaging) exhibited a significant increase in TVB-N levels to 32 ± 0.8 mg, surpassing the acceptable threshold of 25 mg per 100 g of sample. In contrast, even after 14 days, the samples with smart packaging consistently maintained TVB-N levels below the permissible limit, indicating effective preservation (Figure 5f). Figure 5g depicts a radar chart with a comprehensive evaluation of the smart packaging system developed in this work by comparing its performance across multiple criteria against existing packaging reported in the literature. The comparison suggests that the smart packaging in this work performs robustly in several key areas. However, improving the biodegradability and shelf-life extension is still needed.

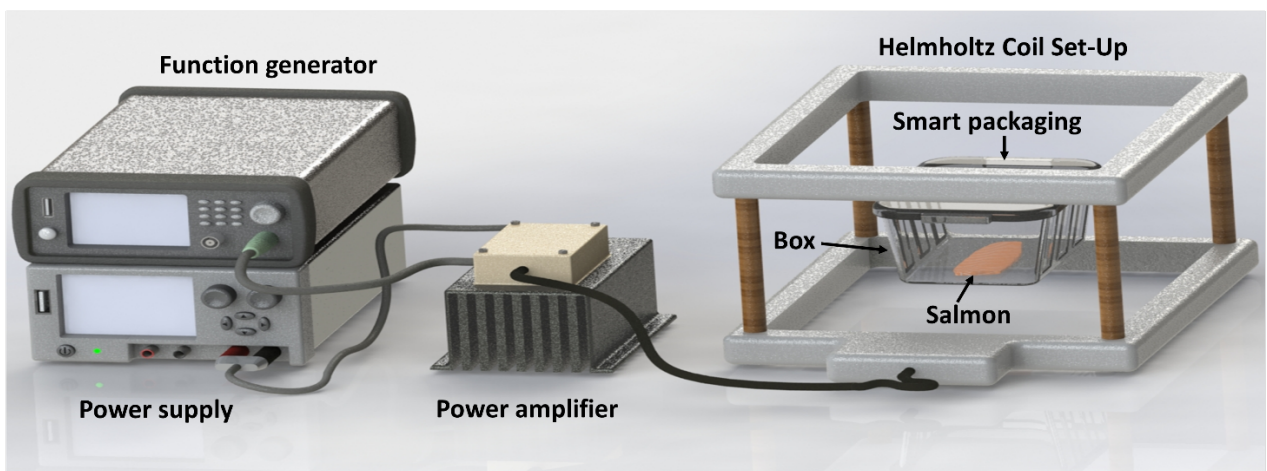

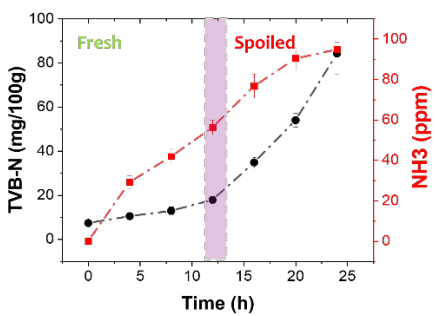
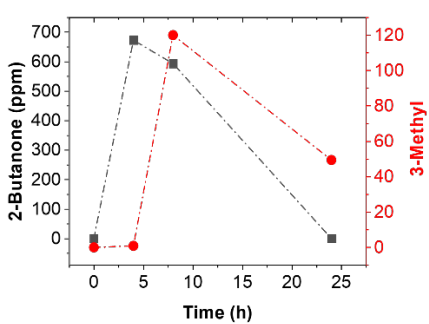
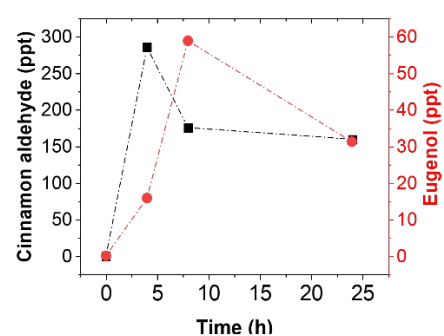
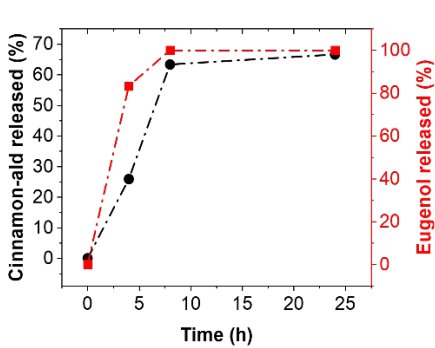
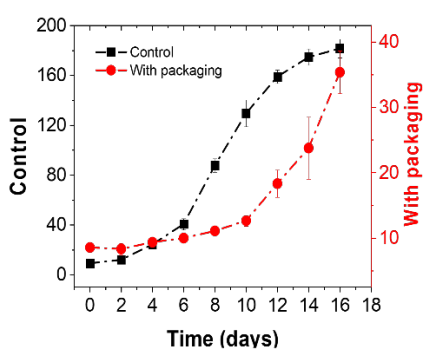
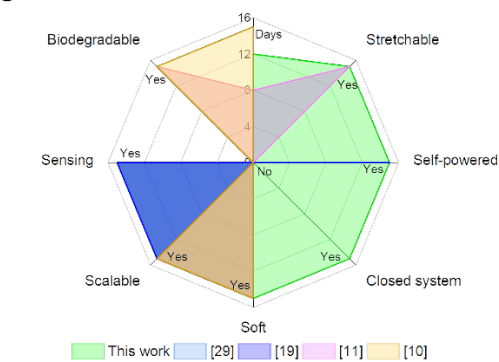

*Figure 5| (a)* the experimental setup used to investigate the performance of the smart packaging with a real sample -salmon-, *(b)*

*humidity and temperature increase inside a box with a salmon, **(c)** TVB-N and NH3 increase inside the box over time, **(e)** 2-Butanone and 3-methyl butanol change over time, **(f)** cinnamaldehyde, and eugenol change over time, **(g)** cumulative release of cinnamaldehyde, and eugenol over a period of 24 h, **(i)** TVB-N increase at 4 °C, **(j)** comparison between this work and the literature.*

In conclusion, our study presents a breakthrough in food packaging with the development of a battery-less, autonomous smart packaging system. This system effectively combines real-time food monitoring and the controlled release of active compounds to significantly extend the shelf life of food products. Employing a gas sensor for freshness monitoring and wirelessly powering the system via an NFC antenna allowed for a controlled release of active compounds, moreover, the system demonstrated the capability to dynamically respond to food spoilage, thus addressing both performance and sustainability challenges in food packaging. The successful extension of food shelf life up to 14 days in our tests underlines the potential of this system as an innovative IoT solution in the food industry, making a substantial contribution to global efforts to reduce food waste and improve sustainability. This research opens new avenues for smart packaging solutions that can adapt and respond to changing food environments, paving the way for a new era of efficiency in food storage and distribution. However, future studies should include the investigation of the exact release mechanism of active compounds.

**Author contributions:**
Conceptualization: AD
Methodology: AD,
Investigation: AD, AM, EL, GW, RR, SI, TNT, ELP
Visualization: AD
Supervision: PL, LP, EB, AA
Writing—original draft: AD
Writing—review & editing: all